\documentstyle[12pt]{article}

\def\+{\;+\;}
\def\-{\;-\;}
\def\*{\,\cdot\,}
\def\p{\mbox{\sf p}}
\def\k{\mbox{\sf k}}
\def\e{\mbox{\sf e}}
\def\E{\mbox{{\large\bf e}}}
\def\Tau{{\cal T}}
\def\one{\mbox{\sf e}_1}
\def\two{\mbox{\sf e}_2}
\def\thr{\mbox{\sf e}_3}
\def\u{\mbox{\sf u}}
\def\w{\mbox{\sf w}}

\def\ds{\displaystyle}

\begin{document}

\title{SOLITONS IN A 3d INTEGRABLE MODEL}

\author{
S. M. Sergeev\\
Branch Institute for Nuclear Physics,\\ 
Protvino 142284, Russia.\\ 
E-mail: sergeev\_ms@mx.ihep.su}

\date{}

\maketitle
\abstract{\scriptsize
Equations of motion for a classical $3d$ discrete model, 
whose auxialiary system is a linear system, are investigated. 
The Lagrangian form of the equations of motion is derived.
The Lagrangian variables are a triplet of ``tau functions''.
The equations of motion for the Triplet of Tau functions
are Three Trilinear equations.
Simple solitons for the trilinear equations are given.
Both the dispersion relation and the phase shift 
reflect the triplet structure of equations.
}

\section{Introduction}

Recently there was proved the exact integrability
of one discrete $3d$ model, classical as well as quantum
\cite{s-3dsympl,s-qem}. This model was formulated originally
in terms of the pure time evolution as the map of the dynamical  
variables from time $t$ to time $t+1$. Namely, for $t$ fixed
the system of the dynamical variables are the system of
\begin{equation}
\ds [\;\u_{\alpha,a,b}\;,\;\w_{\alpha,a,b}\;]\;,\;\;\;
\alpha=1,2,3,\;\;\;a,b\in Z_M\;,
\end{equation}
where $M$ is the finite spatial size of the system,
and the Poisson brackets in the classical case are 
\begin{equation}\label{Poisson}
\ds
\{\u_{\alpha,a,b}\;,\;\w_{\alpha',a',b'}\}\;=\;
\delta_{\alpha,\alpha'}\;\delta_{a,a'}\;\delta_{b,b'}\;
\u_{\alpha,a,b}\,\w_{\alpha,a,b}\;.
\end{equation}
The evolution was formulated as the explicit form of functions $f,g$,
\begin{equation}
\ds
[\u,\w]\;=\;[\u(t),\w(t)]\;\mapsto\;
[\u(t+1),\w(t+1)]\;=\;[f(\u,\w),g(\u,\w)]\;,
\end{equation}
such that the Poisson brackets are conserved by this map.
The existence of the complete set of involutive integrals of motion
was proved in \cite{s-3dsympl,s-qem}.

The map $t\;\mapsto\;t+1$ may be considered as a sort of 
the Hamiltonian 
equations of motion for this classical discrete model. 
In this
paper we will derive their Lagrangian counterpart. Following
\cite{rmk-lybe}, one may guess \`a priori that the Lagrangian
variable is Hirota's tau function and it might obey Hirota's discrete
bilinear equations \cite{Hirota}. 
But we will show that this is not true in
general, although equations we will derive resemble Hirota's ones.

In this note we recall the reader the Hamiltonian form 
of the equations of motion first, then introduce the analogous
of the tau function and derive the Lagrangian form of the 
equations of motion, and finally give the multi-soliton 
solution of them.

\section{Hamiltonian equations of motion}

In this paper we investigate the equations of motion,
so we need no the evolution operator and may choose more
appropriate coordinate system. Actually, the coordinates
we will use here are the light cone frame with respect to previous
$t,a,b$. Let $\one,\two,\thr$ be three translations making
\begin{equation}
\ds\begin{array}{ccl}
\ds\one & : & \ds (t,a,b)\mapsto(t+1,a,b)\;,\\
&&\\
\ds\two & : & \ds (t,a,b)\mapsto(t+1,a+1,b)\;,\\
&&\\
\ds\thr & : & \ds (t,a,b)\mapsto(t+1,a,b+1)\;.
\end{array}\end{equation}
$\one,\two,\thr$ are the elementary orthonormal 
shifts of the three dimensional
cubic lattice. So for any site of the cubic lattice $\p$ given,
the following eight sites form the elementary cube of the lattice:
\begin{equation}\label{cube}
\ds\left(\begin{array}{cccc}
\ds \p\;,&\ds\p+\two+\thr\;,&\ds\p+\one+\thr\;,&\ds\p+\one+\two\;,\\
&&&\\
\ds\p+\one+\two+\thr\;,&\ds\p+\one\;,&\ds\p+\two\;,&\ds\p+\thr\;.
\end{array}\right)\end{equation}

Equations of motion for $\u_{\alpha,\p}$, $\w_{\alpha,\p}$ may
be extracted from the form of $f,g$ in the definition of the 
evolution in \cite{s-3dsympl,s-qem}, and in our coordinates
look like

\noindent
(i)
\begin{equation}\label{ev-1}
\ds\left\{\begin{array}{ccl}
\ds\w_{1,\p+\one} & = & \ds
{\w_{1,\p}\,\w_{2,\p}\,+\,\u_{3,\p}\,\w_{2,\p}\,+\,
\kappa_3\,\u_{3,\p}\,\w_{3,\p}\over\w_{3,\p}}\;,\\
&&\\
\ds\u_{1,\p+\one} & = & \ds
{\kappa_2\,\u_{1,\p}\,\u_{2,\p}\,\w_{2,\p}\over
\kappa_1\,\u_{1,\p}\,\w_{2,\p}\,+\,
\kappa_3\,\u_{2,\p}\,\w_{3,\p}\,+\,
\kappa_1\,\kappa_3\,\u_{1,\p}\,\w_{3,\p}}\;,
\end{array}\right.
\end{equation}
(ii)
\begin{equation}\label{ev-2}
\ds\left\{\begin{array}{ccl}
\ds\w_{2,\p+\two} & = & \ds
{\w_{1,\p}\,\w_{2,\p}\,\w_{3,\p}\over
\w_{1,\p}\,\w_{2,\p}\,+\,
\u_{3,\p}\,\w_{2,\p}\,+\,
\kappa_3\,\u_{3,\p}\,\w_{3,\p}}\;,\\
&&\\
\ds\u_{2,\p+\two} & = & \ds
{\u_{1,\p}\,\u_{2,\p}\,\u_{3,\p}\over
\u_{2,\p}\,\u_{3,\p}\,+\,\u_{2,\p}\,\w_{1,\p}\,+\,
\kappa_1\,\u_{1,\p}\,\w_{1,\p}}\;,
\end{array}\right.\\
\end{equation}
(iii)
\begin{equation}\label{ev-3}
\ds\left\{\begin{array}{ccl}
\ds\w_{3,\p+\thr} & = & \ds
{\kappa_2\,\u_{2,\p}\,\w_{2,\p}\,\w_{3,\p}\over
\kappa_1\,\u_{1,\p}\,\w_{2,\p}\,+\,
\kappa_3\,\u_{2,\p}\,\w_{3,\p}\,+\,
\kappa_1\,\kappa_3\,\u_{1,\p}\,\w_{3,\p}}\;,\\
&&\\
\ds\u_{3,\p+\thr} & = & \ds
{\u_{2,\p}\,\u_{3,\p}\,+\,
\u_{2,\p}\,\w_{1,\p}\,+\,
\kappa_1\,\u_{1,\p}\,\w_{1,\p}\over \u_{1,\p}}\;.
\end{array}\right.
\end{equation}

Describe in a couple of words the idea of the derivation of
equations (\ref{ev-1}-\ref{ev-3}).
These relations may be obtained as the zero curvature condition
of the following system. Let $\varphi_{\p}$ be an auxiliary
variable assigned to the sites of the cubic lattice.
Consider three orthogonal plaquettes of the cube endowed by the
following
relations
\begin{equation}\label{linear}
\ds\begin{array}{ccc}
\ds 0\;=\;f_{1,\p} &\stackrel{def}{=}& \ds
\varphi_{\p}\;\w_{1,\p}\;-\;
\varphi_{\p+\two+\thr}\;\u_{1,\p}\;+\;
\varphi_{\p+\two}\;+\;
\varphi_{\p+\thr}\;\kappa_1\;\u_{1,\p}\;\w_{1,\p}\;,\\
&&\\
\ds 0\;=\;f_{2,\p} &\stackrel{def}{=}& \ds
\varphi_{\p}\;+\;
\varphi_{\p+\one+\thr}\;\kappa_2\;\u_{2,\p}\;\w_{2,\p}\;+\;
\varphi_{\p+\one}\;\w_{2,\p}\;-\;
\varphi_{\p+\thr}\;\u_{2,\p}\;,\\
&&\\
\ds 0\;=\;f_{3,\p} &\stackrel{def}{=}& \ds
-\;\varphi_{\p}\;\u_{3,\p}\;+\;
\varphi_{\p+\one+\two}\;\w_{3,\p}\;+\;
\varphi_{\p+\two}\;+\;
\varphi_{\p+\one}\;\kappa_3\;\u_{3,\p}\;\w_{3,\p}\;.
\end{array}\end{equation}
Easy to see that the number of linear equations is
three times greater then the number of $\varphi_{\p}$.
This means that the coefficients of the system of linear 
equations must obey extra conditions. 
As an example consider
a cube. Six plaquettes give six linear relations 
for eight corner $\varphi_{\p}$:
\begin{equation}
\ds f_{\alpha,\p}\;=\;f_{\alpha,\p+\e_\alpha}\;=\;0\;,\;\;\;
\alpha\;=\;1,2,3\;.
\end{equation}
Demand that between these
six relations there are only four linearly independent,
so that the cube relations fix only four of eight corner
linear variables $\varphi_{\p}$
(e.g. we just may express
$\varphi_{\p+\one+\two}$, $\varphi_{\p+\two+\thr}$, 
$\varphi_{\p+\one+\thr}$ and $\varphi_{\p+\one+\two+\thr}$
via independent $\varphi_{\p}$, $\varphi_{\p+\one}$,
$\varphi_{\p+\two}$ and $\varphi_{\p+\thr}$ and nothing more).
This demand gives equations of motion (\ref{ev-1}-\ref{ev-3})
immediately.

\section{Lagrangian equations of motion}

Three simple relations between the equations of motion are to be
mentioned:
\begin{equation}\label{simple}
\ds\left\{\begin{array}{ccc}
\ds\w_{1,\p}\,\w_{2,\p} & = & \ds \w_{1,\p+\one}\,\w_{2,\p+\two}\;,\\
&&\\
\ds\u_{2,\p}\,\u_{3,\p} & = & \ds \u_{2,\p+\two}\,\u_{3,\p+\thr}\;,\\
&&\\
\ds\u_{1,\p}/\w_{3,\p} & = & \ds \u_{1,\p+\one}/\w_{3,\p+\thr}\;.
\end{array}\right.\end{equation}
Following \cite{rmk-lybe}, 
a parametrization of $\u,\w$ in terms of tau functions
must turn relations (\ref{simple}) into tautologies.
Thus without lost of generality (we deal with the infinite system,
avoiding hence all the boundary problems)
\begin{equation}\label{tau-param}
\ds\left\{\begin{array}{ll}
\ds\w_{1,\p}^{}\;=\;
{\Tau_{3,\p+\two}\over\Tau_{3,\p}}\;,\;\;\;\;&
\ds\u_{1,\p}^{}\;=\;
{\Tau_{2,\p}\over\Tau_{2,\p+\thr}}\;,\\
&\\
\ds\w_{2,\p}^{}\;=\;
{\Tau_{3,\p}\over\Tau_{3,\p+\one}}\;,&
\ds\u_{2,\p}^{}\;=\;
{\Tau_{1,\p}\over\Tau_{1,\p+\thr}}\;,\\
&\\
\ds\w_{3,\p}^{}\;=\;
{\Tau_{2,\p}\over\Tau_{2,\p+\one}}\;,&
\ds\u_{3,\p}^{}\;=\;
{\Tau_{1,\p+\two}\over\Tau_{1,\p}}\;.
\end{array}\right.\end{equation}
As usual, $\Tau_{\alpha,\p}$ may contain a quadratic and linear 
pre-exponent,
\begin{equation}\label{tau-pure}
\ds\Tau_{\alpha,\p}\;=\;
\E^{{1\over 2}\;(\p ,Q_\alpha,\p)\;+\;
(\mbox{\sf q}_\alpha,\p)}\,\cdot\,\tau_{\alpha,\p}\;,
\end{equation}
where $\mbox{\sf q}_\alpha$ and diagonals of 
$Q_\alpha$ are arbitrary, but all non-diagonal
elements of $Q_\alpha$ must coincide (in the natural basis
$\e_\alpha$).

Substituting now the parametrization
(\ref{tau-param}) to (\ref{ev-1}-\ref{ev-3})
and taking (\ref{tau-pure}) into account,
we obtain Three Trilinear relations for Three $\tau$
functions:
\noindent
\begin{equation}\label{tril-1}
\ds\left\{\begin{array}{ccc}
\ds r_1^{}\;\tau_{1,\p+\two+\thr}\;\tau_{2,\p}\;\tau_{3,\p} & = & 
\ds \tau_{1,\p}\;\tau_{2,\p+\thr}\;\tau_{3,\p+\two}\\
&&\\&+&
\ds s_2^{}\;\tau_{1,\p+\two}\;\tau_{2,\p+\thr}\;\tau_{3,\p}\\
&&\\&+&
\ds s_3^{-1}\;\tau_{1,\p+\thr}\;\tau_{2,\p}\;\tau_{3,\p+\two}\;,
\end{array}\right.\end{equation}
\begin{equation}\label{tril-2}
\ds\left\{\begin{array}{ccl}
\ds r_2^{}\;\tau_{1,\p}\,\tau_{2,\p+\one+\thr}\,\tau_{3,\p} & = & 
\ds \tau_{1,\p+\thr}\,\tau_{2,\p}\,\tau_{3,\p+\one}\\
&&\\&+&
\ds s_3^{}\;\tau_{1,\p}\,\tau_{2,\p+\thr}\,\tau_{3,\p+\one}\\
&&\\&+&
\ds s_1^{-1}\;\tau_{1,\p+\thr}\,\tau_{2,\p+\one}\,\tau_{3,\p}\;,
\end{array}\right.\end{equation}
\begin{equation}\label{tril-3}
\ds\left\{\begin{array}{ccl}
\ds r_3^{}\;\tau_{1,\p}\,\tau_{2,\p}\,\tau_{3,\p+\one+\two} & = & 
\ds \tau_{1,\p+\two}\,\tau_{2,\p+\one}\,\tau_{3,\p}\\
&&\\&+&\ds
s_1^{}\;\tau_{1,\p+\two}\,\tau_{2,\p}\,\tau_{3,\p+\one}\\
&&\\&+&
\ds s_2^{-1}\;\tau_{1,\p}\,\tau_{2,\p+\one}\,\tau_{3,\p+\two}\;.
\end{array}\right.\end{equation}
Here we gather all $\kappa$-s and all the pre-exponents
into arbitrary $s_\alpha,r_\alpha$, $\alpha=1,2,3$. 
One may change them in any appropriate way.

Because of the possibility to introduce the Poisson brackets 
(\ref{Poisson}) in the system of $\u_{\alpha,\p},\w_{\alpha,\p}$,
the original equations of motion (\ref{ev-1}-\ref{ev-3})
may be regarded as a kind of the Hamiltonian equations of motion.
Contrary to that, the tautological parametrization of (\ref{simple})
resembles the introduction of a velocity instead of a momentum,
so that the number of variables decreases twice, so the trilinear 
relations (\ref{tril-1}-\ref{tril-3}) are nothing but the 
Lagrangian form
of the equations of motion.

(\ref{tril-1}-\ref{tril-3}) may be reduced to
Hirota's bilinear discrete equation in the limit
\begin{equation}
\ds\kappa_1\;<<\;\kappa_2\;=\;\kappa_3\;<<\;1\;.
\end{equation}

\section{Solitons}

Now describe the simple solitons of the trilinear
relations (\ref{tril-1}-\ref{tril-3}). 

Let $(\alpha,\beta,\gamma)$ be any cyclic permutation of $(1,2,3)$.
Fix $r_\alpha$ via
\begin{equation}
\ds
r_\alpha\;=\;1\;+\;s_\beta\;+\;{1\over s_\gamma}\;.
\end{equation}
Then let the exponent
\begin{equation}
\ds W\;=\;\E^{i\,(\k,\p)}\;=\;\E^{\ds
i\,k_1\,p_1+i\,k_2\,p_2+i\,k_3\,p_3}
\end{equation}
for $s_\alpha$ given, is defined by
\begin{equation}\label{spectr}
\ds
\E^{\ds i\;k_\alpha}\;=\;
{\lambda_\alpha\;((\lambda_\alpha-\lambda_\gamma)\;+\;
s_\alpha\,(\lambda_\alpha-\lambda_\beta))\over
\lambda_\beta\,(\lambda_\alpha-\lambda_\gamma)\;+\;
s_\alpha\,\lambda_\gamma\,(\lambda_\alpha-\lambda_\beta)}\;.
\end{equation}
This parametrization
of the dispersion relation we will exhibit as $W=W(\lambda)$. 
For the dispersion curve $W=W(\lambda)$ given, let for the shortness
\begin{equation}
\ds
W^{(k)}_\alpha\;=\;\lambda_\alpha^{(k)}\;W(\lambda^{(k)})\;.
\end{equation}
The phase shift $D$ is defined by
\begin{equation}
\ds
D(\lambda,\lambda^\prime)\;=\;
{d(\lambda,\lambda^\prime)\,\cdot\,d(\lambda^{-1},\lambda^{\prime-1})
\over
d(\lambda^{-1},\lambda^\prime)\,\cdot\,d(\lambda,\lambda^{\prime-1})}
\;,
\end{equation}
where
\begin{equation}
\ds
d(\lambda,\lambda^\prime)\;=\;\det\;\left(
\begin{array}{ccc}
\ds 1 &\ds 1 &\ds 1 \\
&&\\
\ds \lambda_1^{} & \ds \lambda_2^{} & \ds \lambda_3^{}\\
&&\\
\ds \lambda_1' & \ds \lambda_2' & \ds \lambda_3'
\end{array}\right)\;,
\end{equation}
and $\lambda^{-1}$ stands for 
$\{\lambda_1^{-1},\lambda_2^{-1},\lambda_3^{-1}\}$.
For the shortness let
\begin{equation}
\ds D^{(k,l)}\;=\;D(\lambda^{(k)},\lambda^{(l)})\;.
\end{equation}
One soliton solution is
$\ds\tau_{\alpha,\p}^{(1)}\;=\;1\;-\;W_\alpha$.
Two solitons are 
\begin{equation}
\ds\tau_{\alpha,\p}^{(2)}\;=\;
1\;-\;W_\alpha^{(1)}\;-\;W_\alpha^{(2)}\;+\;
D^{(1,2)}\,W_\alpha^{(1)}\,W_\alpha^{(2)}\;.
\end{equation}
Three solitons are
\begin{equation}
\ds\begin{array}{ccl}
\ds\tau_{\alpha,\p}^{(3)} & = & \ds
1\;-\;W_\alpha^{(1)}\;-\;W_\alpha^{(2)}\;-\;W_\alpha^{(3)}\\
&&\\
&&\ds
+\;D^{(1,2)}\,W_\alpha^{(1)}\,W_\alpha^{(2)}
+D^{(1,3)}\,W_\alpha^{(1)}\,W_\alpha^{(3)}
+D^{(2,3)}\,W_\alpha^{(2)}\,W_\alpha^{(3)}\\
&&\\
&&-\;
D^{(1,2)}\,D^{(1,3)}\,D^{(2,3)}\,
W_\alpha^{(1)}\,W_\alpha^{(2)}\,W_\alpha^{(3)}\;.
\end{array}\end{equation}
And so on, common expression is usual.
Remarkable is that the dispersion relation
for the exponents $\E^{i\,k_\alpha}$ may be parametrized by
their origin $\lambda_\alpha$, (\ref{spectr}),
and parameters $s_\alpha$ are defined by an arbitrary linear
pre-exponent.

\section{Discussion}

System (\ref{tril-1}-\ref{tril-3}) contains a set of bilinear
relations
as its necessary conditions. Looking for a solution of trilinear
relations in a form of holomorphic functions, we get the following
set of necessary relations
\begin{equation}\label{holomorphic}
\Lambda_{\alpha,\p+\e_\alpha}\,\tau_{\alpha,\p}\;=\;
\tau_{\beta,\p+\e_\alpha}\,\tau_{\gamma,\p}\;+\;
s_\alpha\;\tau_{\beta,\p}\,\tau_{\gamma,\p+\e_\alpha}\;,
\end{equation}
where as before $(\alpha,\beta,\gamma)$ are any cyclic permutation
of $(1,2,3)$, and auxiliary functions
$\Lambda_{\alpha,\p}$ are supposed to be holomorphic. Thus 
(\ref{tril-1}-\ref{tril-2}) have a host of solutions similar
to that of Hirota's equation. Namely, identifying (\ref{holomorphic})
with a set of Fay's or Pl\"ucker's relations, we may obtain
elliptic or determinant solution of (\ref{tril-1}-\ref{tril-3}).
Probably, the exception is a Bethe -- ansatz -- type solution of 
(\ref{holomorphic}), because in this case (\ref{tril-1}-\ref{tril-3})
do not follow from (\ref{holomorphic}) tautologically.
Because of all these do not contain anything so surprising as
the solitons, we do not give any explicit formula here. 
Mention only two aspects.

First one is the symmetry group of (\ref{tril-1}-\ref{tril-3}).
For Hirota's equation all the symmetries with respect to 
permutations and reflections of the space directions are obvious.
In the case of (\ref{tril-1}-\ref{tril-3}) the Cube group is less
trivial because of it acts on the parameters $s_\alpha,r_\alpha$.
As an example give the realisation of the complete reflection $P$.
For a solution $\tau_{\alpha,\p}$ of the trilinear equations with
some $s_\alpha,r_\alpha$ given let 
\begin{equation}
\overline{\tau}_{\alpha,\p}\;=\;\tau_{\alpha,\p-\e_\alpha}\;,\;\;
\overline{s}_\alpha\;=\;
{r_\beta\over r_\gamma}\;{s_\alpha\over s_\beta\,s_\gamma}\;,\;\;
\overline{r}_\alpha\;=\;
{s_\beta\over s_\gamma}\;{r_\beta\,r_\gamma\over r_\alpha}\;.
\end{equation}
Then the set of ``overlined'' objects obey the 
same set of the trilinear relations (\ref{tril-1}-\ref{tril-3})
but with reflected $\e_\alpha\;\mapsto\;-\;\e_\alpha$ for all
$\alpha$.

Mention also the possibility to construct the complete solution
of original finite evolution model in terms of the elliptic functions 
on a finite genus curve.
In such case one may consider a non-homogeneous and non-equidistant
version of the trilinear relations with a boundary conditions
taken into account. The number of the parameters
(moduli, initial divisor and homogeneousless data) will coincide
with the number of variables of the initial state. 
This will be the subject of a separate paper.

\noindent
{\bf Acknowledgments.} I would like to thank Rinat Kashaev 
for wery useful explanaions and comments.
The work was supported by the RFBR grant No. 98-01-00070.

\end{document}